\documentclass{article}

\usepackage[utf8]{inputenc}
\usepackage{amsmath,amsthm}
\usepackage{amsfonts}
\usepackage{mathtools}
\usepackage{xcolor}
\usepackage{enumitem}
\usepackage{array}
\usepackage{url}

\newtheorem{proposition}{Proposition}
\newtheorem{observation}{Observation}

\newtheorem{corollary}{Corollary}

\newtheorem{theorem}{Theorem}
\theoremstyle{definition}
\newtheorem{remark}{Remark}
\newtheorem{definition}{Definition}

\newtheorem{example}{Example}

\newcommand{\calC}{\mathcal{C}}
\newcommand{\calCN}{\mathcal{CN}}

\title{Egalitarian and Just Digital Currency Networks}

\author{Gal Shahaf \\
Weizmann Institute \\
\and
Ehud Shapiro \\
Weizmann Institute \\
\and
Nimrod Talmon \\
Ben-Gurion University \\
}

\date{}
\begin{document}

\pagestyle{plain}

\maketitle

\begin{abstract}
Cryptocurrencies are a digital medium of exchange with decentralized control that renders the community operating the cryptocurrency its sovereign.  Leading cryptocurrencies use proof-of-work or proof-of-stake to reach consensus, thus are inherently plutocratic. This plutocracy is reflected not only in control over execution, but also in the distribution of new wealth, giving rise to ``rich get richer'' phenomena.
Here, we explore the possibility of an alternative digital currency that is egalitarian in control and just in the distribution of created wealth. Such currencies can form and grow in grassroots and sybil-resilient way.
A single currency community can achieve distributive justice by \emph{egalitarian coin minting}, whereby each member mints one coin at every time step. Egalitarian minting results, in the limit, in the dilution of any inherited assets and in each member having an equal share of the minted currency, adjusted by the relative productivity of the members. Our main theorem shows that a currency network, where agents can be members of more than one currency community, can achieve distributive justice globally across the network by \emph{joint egalitarian minting}, whereby each agent mints one coin in only one community at each timestep. Specifically, we show that a sufficiently large intersection between two communities -- relative to the gap in their productivity -- will cause the exchange rates between their currencies to converge to 1:1, resulting in global distributive justice.
%
\end{abstract}

\section{Introduction}

Money is nothing but a piece of paper; or a string of bits, perhaps. In modern history, fiat money is issued and controlled by rulers and governments.
Following Bitcoin~\cite{bitcoin}, many blockchain-based cryptocurrencies were introduced~\cite{cryptocurrencypaper}. Their technology and distributed protocol renders the community operating the currency its sovereign as, unlike in standard computer systems, there is no third party that may exert control over the system, e.g., shut it down.

In existing cryptocurrencies, however, most control and benefit lies in the hands of the few\footnote{See, e.g., \url{https://bitcoinera.app/arewedecentralizedyet/}.} -- their founders, early adopters, and large stakeholders (e.g. large ``mining pools'')~\cite{miningpools}.
In this paper we explore the possibility of forming an egalitarian and just digital currency that may form currency networks in a grassroots manner. Our goal here is the design of a digital currency that may be issued by all, where both control and benefit are distributed in an egalitarian way among the people participating in the creation and use of the currency. Such a currency implements distributive justice in the sense that each person enjoys an equal share of the created currency.

One key challenge in this task is the presence of fake and duplicate identities, aka \textit{sybils}, that may be employed by their operators in order to tilt control and wealth in their favor. Indeed, justice and equality can be achieved only if the parties to the currency are genuine (unique and singular) agents of the participating people~\cite{gid}, thus excluding sybils.  
We first observe that sybils cannot penetrate small communities of people that know and trust each other and that, indeed, trust communities can grow in a sybil-resilient way by employing graph-based properties~\cite{csr} of genuine identifiers~\cite{gid}, using various mechanisms as admission rules to the community~\cite{sutherland2002controlled}, or utilizing some machine learning algorithms~\cite{gong2014sybilbelief}.

However, as we wish our medium to be scalable, our further goal is to build this digital currency in a grassroots way. In particular, our paper may be viewed as means for a joint, safe scale-up of such communities, concentrating on the aspect of distributive justice as we rely on the infrastructure of digital social contracts~\cite{dsc} for equality in execution, and techniques such as mutual sureties~\cite{gid} and sybil-resilient community expansion~\cite{csr} for sybil resilience. 

The key, high-level differences between our proposed digital currency and most existing cryptocurrencies are outlined below:
\begin{itemize}

\item \textbf{Equality:}
Leading cryptocurrencies employ either proof-of-work (PoW) or proof-of-stake (PoS) systems~\cite{cryptocurrencies}. As such, they are inherently plutocratic, since control over the behavior of the system is positively correlated with the computing power or amount of currency available to different parties. A cryptocurrency is \emph{egalitarian} if control over the execution and modification of the currency system is shared equally among the parties to the currency.
Such equality can be guaranteed using digital social contracts~\cite{dsc} over genuine identifiers~\cite{gid}.

\item \textbf{Distributive justice:}
Leading cryptocurrencies do not aim for justice, distributive or otherwise. Newly minted coins are allocated to parties with superior computing power (PoW) or larger amounts of currency (PoS).
A cryptocurrency satisfies \emph{distributive justice} if each agent enjoys an equal share of the newly created value of the currency.
Here, we formally define distributive justice in this context and spell out conditions that give rise to it.
In particular, a single currency community can achieve distributive justice by egalitarian coin minting, where each member mints one coin in every time step. Assuming the community has only genuine members and no sybils, egalitarian minting results, in the limit, in the dilution of any inherited assets and in each member having an equal share of the minted currency, adjusted by the relative productivity of the members. In a currency network, where people can be members of more than one currency community, a joint egalitarian minting regime in which each person mints one coin in only one community in each timestep, allows market forces to achieve distributive justice globally across the network, under conditions that we discuss.

\item \textbf{Grassroots Sybil-Resilience:}
Leading cryptocurrencies are monolithic, in that there is one community using the cryptocurrency (e.g., one blockchain in which the bitcoin transactions are recorded). Here, we aim at a grassroots architecture that allows currency communities to form independently, allowing people from different communities to trade and exchange their currencies, and eventually form a currency network that serves as a joint, grassroots medium of exchange.  Our method is sybil-resilient in that sybils in a currency network affect only the currency communities that harbour them. We defer the discussion on the specific sybil resilience achieved in such networks for future work.

\end{itemize}

In this paper, we first begin with a single currency and provide a formal definition for a just distribution among its agents. Intuitively, distributive justice is satisfied if every member of the currency community is granted, initially, an equal share of the currency, and may trade its portion as it pleases. Formally, at every time step, the diluted balance of every agent amounts to its equal share plus its diluted cashflow up to this point.
We then present a richer notion of \textit{asymptotic justice}, where distributive justice is reached in the limit. With this notion, distributive justice can be reached even if agents begin with different initial amounts of the currency; as such it models distributive justice in the face of unequal inheritances.
To achieve asymptotic justice, the difference between the diluted balance and cashflow must converge to an equal share of the currency, but these quantities need not match at all times.  We show that this notion of justice may be realized via egalitarian coin minting, which provides a form of  \textit{Universal Basic Income} (UBI). That is, a community in which each member mints an equal amount of coins in every time step results in asymptotic justice, regardless of the initial balance of the agents and the differences in the time of of joining the community.


Envisioning the emergence of different and independent currency communities, each employing their own egalitarian minting regime as describe above, we then analyze conditions under which multiple communities may inter-operate in such a way that, jointly, genuine agents in all communities will get an equal share of the joint created value of all currencies; that is, we set to investigate the possibility of achieving global distributive justice in a situation where many independent currencies are used at once.
To this end, we define the notion of a \emph{currency network}, in which several currency communities operate simultaneously. The formal definition of a currency network is given below; in essence, it is a tuple of communities that employ independent currencies (each coin belongs to a single currency). The network structure arises from chain payments via agents that are members in multiple communities simultaneously. This model is a direct generalization of \textit{credit networks}~\cite{creditnetworks,ghosh2007mechanism,ramseyer2019constrained,dandekar2011liquidity,dandekar2015strategic}.

In order to analyze the dynamics of such networks and the economic consequences of such dynamics, we apply the \textit{free exchange economy} model~\cite{moore2007pure} for the emergence of exchange rates among the different currencies. Based on these rates, we extend the definition of distributive justice to a currency network, and provide sufficient conditions under which distributive justice is satisfied. Importantly, these conditions rely on the currency volumes being in perfect balance with the marginal rates of substitution among the currencies. This balance requires calibration with every alternation in the network structure (i.e., the admission of a new member, etc.), and is thus hard to maintain without a frictionless and efficient trade among the currencies.

With these assumptions, we extend the notion of asymptotic justice to currency networks. Our main result in this setting provides sufficient conditions under which asymptotic justice is achieved under an \textit{egalitarian minting regime}. 
That is, in order to obtain distributive justice in the limit, the substantial collaboration among the different communities is expressed in jointly ensuring that every agent may mint one coin of only one currency at every time step. Agents may choose which coin to mint from the different currency communities in which they are members.
Specifically, our main result shows that exchange rates between two communities will converge to 1:1 and asymptotic justice would follow, as long as the following conditions hold:
\begin{enumerate}
\item Agents behave myopically, in that each agent mints the highest valued coin at every time step;
\item The network is efficient, in that agents trade coins in order to maximize their utilities, causing equilibria to be reached infinitely often; 
\item The intersections among the two communities is sufficiently large to compensate for the productivity gap between them.
\end{enumerate}
Our focus in this paper is on the economic analysis of currency networks, as described above. Ultimately, we aim to implement such currencies using digital social contracts, and show social contract schemes for single- and multi-currency egalitarian minting~\cite{dsc}.  Our analysis shows how distributive justice can be achieved globally in a network of egalitarian and grassroots digital currencies. Importantly, while a distributed implementation of this model must deal with the asynchronous nature of the underlying communication network among the agents (as in digital social contracts~\cite{dsc}), here, for simplicity, we assume a synchronous model of computation.

Finally, we discuss the relation between people and their agents, and show that if a currency community is genuine then it can achieve distributive justice among its owners.  In a currency network with a genuine (sybil-free) subnet, distributive justice can be achieved among all owners of the subnet.

\subsection{Organization}

After reviewing related work, we proceed with the notion of a single currency community at Section \ref{section:singlecurrency}, where we define initial and asymptotic distributive justice, and discuss means for achieving them. We then address currency networks at Section \ref{section:currencynetworks}, where we discuss the emergence of exchange rates via the free exchange economy model and extend the definitions of justice to this richer setting. Then, at Section~\ref{section:justice}, we analyze sufficient conditions for asymptotic justice in a network under an egalitarian minting regime.

\subsection{Related Work}

Mathematically, the main predecessor for personal currency networks are credit networks~\cite{dandekar2011liquidity,dandekar2015strategic,ramseyer2019constrained,defigueiredo2005trustdavis,karlan2009trust,ghosh2007mechanism}, 
and some of the results and analyses of credit networks carry over to personal currency networks.
The key difference between credit networks and our newly proposed digital currency networks is that credit networks \underline{assume} the existence of an objective measure of value, namely, an outside currency, whereas currency networks aim to \underline{create} an objective measure of value.

While credit networks inspired some cryptocurrencies, including Ripple~\cite{schwartz2014ripple} and Stellar~\cite{mazieres2015stellar}, they all had to choose an external currency to peg credit to:
  Ripple has chosen to provide its own cryptocurrency, XRP, the production of which is controlled by the Ripple Foundation (who owns the majority of minted XRP coins), while Stellar chose to be a ``stablecoin'', pegging the credit to a basket of fiat currencies.

Practically, the most related cryptocurrencies are the trust-based currencies of Circles~\cite{circles} and Duniter~\cite{duniter}. Both create money through Universal Basic Income (UBI) to their members.  Circles is a smart contract on top of Ethereum and is still a concept under development. Duniter is a cryptocurrency with an active community of mostly-French users; it anticipated the idea of egalitarian coin minting presented here and has a mechanism of sybil-resilience, being an indication that the conceptual and mathematical framework presented here may be viable.

A UBI-based currency community is a possibility, as demonstrated by Duniter, and is consistent with our mathematical model. Here, in particular, we study joint-UBI regimes, supporting the grassroots formation of multiple currencies; so we do not only concentrate on a single currency community (like Duniter and Circles), but anticipate a network, consisting of many such currencies, and study their joint economic behavior.
Indeed, Duniter is not grassroots in the sense that it does not provide conceptual or architectural foundation for multiple independent Duniter-like currency communities to form and interoperate, like we do.
  
\section{A Currency Community}\label{section:singlecurrency}

Here we first describe a cryptocurrency community that is equal and just, provided it is sybil-free. 
We expect people to participate in a currency community via computational agents, and assume a one-to-one correspondence between people and the agents and refer to the computational agents as ``it''.
Hence, 
Such a sybil-free community may be simply a small-scale community in which all agents know and trust each other, or a larger-scale community that grows in a sybil-resilient way~\cite{gid,csr}.
We first define such a currency community formally, and analyze economic properties of its dynamics, showing in particular that distributive justice can be achieved in the limit using an egalitarian minting regime in which each agent mints a single coin in each timestep.  A digital social contract that implements egalitarian execution of a currency community and egalitarian minting is described elsewhere~\cite{dsc}.

\begin{definition}[Currency Community]
A \textit{Currency Community} is a tuple $\calC = (V,C,h)$, where $V$ is a set of agents, $C$ is a set of fungible coins, and $h:C\xrightarrow{} V$ is a configuration function that indicates the \textit{holder} of each coin $h(c)\in V$. 
\end{definition}
Coins are fungible in the sense defined below.
We shall also use the inverse function $h^{-1}(v):=\{c\in C~|~ h(c) = v\}$ to denote the coins \textit{held} by agent $v\in V$.

We regard the currency as a medium of exchange for goods and services. The fundamental operation in a currency is a \emph{payment}, i.e., the transfer of a coin from a payer to a payee.

\begin{definition}[Payment]\label{definition:currency-payment}
Let $\calC=(V,C,h)$ be a currency community and let $u,v \in V$. A \emph{payment} from $u$ to $v$ is a transfer from $u$ to $v$ of a coin $c\in C$, initially held by $u$. The result of such a payment, denoted by $\calC \xrightarrow{pay(c,u,v)}\calC'$, is the currency community $\calC'=(V,C,h')$, in which:

$$h'(x) := \begin{cases} 
v & \mbox{if } x = c\ ,  \\ 
h(x) &  otherwise\ . \end{cases}
$$
\end{definition}


We observe that payments are reversible.

\begin{observation}[Reversibility in a Single Currency]\label{ob: rev single currency}
If $\calC$ is a currency community and $\calC\xrightarrow{pay(c,u,v)}\calC'$, then
$\calC'\xrightarrow{pay(c,v,u)}\calC$.
\end{observation}

\begin{proof}
By Definition \ref{definition:currency-payment}, exchanging a coin back and forth results in the initial configuration.
\end{proof}

\subsection{A Currency Community History}

We wish to better understand the economic properties of a currency community, in particular, to explore the possibility of achieving distributive justice within the community. To this end, and since we envision a digital currency built with the currency community model in its core, we take the following approach:
  As the economy of a currency community takes place in a dynamic setting, where agents trade coins with each other for goods and services, we consider currency community dynamics.

We assume a dynamic setting with discrete time steps, where coins may be minted periodically by the agents. We mention that this can be implemented by a digital social contract~\cite{dsc} among the participants.  We note that while the formal model digital social contracts, as well as any feasible realization of it,  are asynchronous, we nevertheless assume a synchronous setting as a simpler first step, in particular a notion of time is needed for egalitarian coin minting.

\begin{definition}[Currency Community History]\label{def: community history}
A \emph{currency network history} is a sequence of currency communities $\calC_0, \calC_1, \calC_2, \ldots$,
$\calC_t=(V_t,C_t,h_t)$, $t > 0$, with the following monotonic attributes:
\begin{itemize}
    \item \textit{Agent growth}:  $V_t\subseteq V_{t+1}$ for all $t\geq 0$.
    \item \textit{Coin growth}:  $C_t\subseteq C_{t+1}$ for all $t\geq 0$.
\end{itemize}

\end{definition}

That is, intuitively, we assume that the coin configuration may vary and that new agents and new coins may be added over time.  We leave natural extensions and generalizations of these dynamics (i.e., to accommodate agent departures, coin burns, etc.) for future research.

For the analysis of currency community histories, we employ the notation $V:=\bigcup_t V_t$ to denote all agents throughout history, and define the following.

\begin{definition}[Balance, Income, Revenues and Expenses]
Let $\calC_0, \calC_1, \calC_2, \ldots$. denote a currency community history. Then, we define the following:
\begin{itemize}

\item \textbf{Balance:} The \textit{balance} of agent $v$ at time $t$ is the number of coins held by $v$ at that time, denoted by:
$$b_t(v) = |h_t^{-1}(v)|\ .$$

\item \textbf{Income:} The \textit{income} of agent $v$ at time $t$ is the number of newly minted coins held by $v$, denoted by:

$$ m_t(v) = | h_{t}^{-1}(v) \cap (C_t \setminus C_{t-1}) |\ .$$

\item \textbf{Revenue:} The \textit{revenue} of agent $v$ at time $t$ is the number of coins in $C_{t-1}$ that were added to $v$'s account due to trade, denoted by:

$$ rev_t(v) = | (h_{t}^{-1}(v) \cap C_{t-1} ) \setminus h_{t-1}^{-1}(v) |\ .$$

\item \textbf{Expenses:} The \textit{expenses} of agent $v$ at time $t$ are the number of coins subtracted from $v$'s account due to trade, denoted by:

$$ exp_t(v) = | h_{t-1}^{-1}(v) \setminus h_{t}^{-1}(v)| \ .$$

\end{itemize}
\end{definition}

The relations between these notions are formally expressed in the following:

\begin{observation}
    For every $t>0$ we have
    \begin{equation}\label{eq: bal difference}
    m_t(v) + rev_t(v) - exp_t(v) = b_t(v) - b_{t-1}(v)\ .
    \end{equation}
    
\end{observation}

\begin{proof}
    As $h_{t-1}^{-1}(v)\subseteq C_{t-1}$, we have 
    \begin{align*}
        m_t(v) + rev_t(v) &= | h_{t}^{-1}(v) \cap (C_t \setminus C_{t-1}) |\\
        &+ | (h_{t}^{-1}(v) \cap C_{t-1} ) \setminus h_{t-1}^{-1}(v) |\\
        & = | h_{t}^{-1}(v) \setminus h_{t-1}^{-1}(v) |\ .
    \end{align*}
    
    It follows that $m_t(v) + rev_t(v) - exp_t(v)$ equals
    \begin{align*}
        |h_{t}^{-1}(v) \setminus h_{t-1}^{-1}(v)| &- | h_{t-1}^{-1}(v) \setminus h_{t}^{-1}(v)|\\
        &= |h_{t}^{-1}(v)| - |h_{t-1}^{-1}(v)|\\
        &= b_t(v) - b_{t-1}(v)\ ,
    \end{align*}
which finishes the proof.
\end{proof}

Summing up, we conclude the following:

\begin{corollary}
    For every $t>0$ we have
    \begin{equation}\label{eq: bal sum}
    b_t(v) = b_0(v) + \sum_{s=1}^t \big(m_t(v) + rev_t(v) - exp_t(v) \big)\ .
    \end{equation}
\end{corollary}

That is, the balance of an agent equals its initial endowment plus its income and cash-flow up to this point. 

\subsection{Justice in a Single Currency}

Given the above definitions and observations, we are ready to formally define our desired property of distributive justice, in which, intuitively, every agent is granted an equal share of the currency value. We then demonstrate monetary regimes which realize distributive justice. The fundamental definition of a just currency is the following:  

\begin{definition}[Distributive Justice] \label{def: dist. justice}
    A currency community history is said to be \textit{just} if for every $t\geq 0$ and $v\in V_t$:
    
    $$\frac{b_t(v)}{|C_t|} - \frac{\sum_{s=1}^{t} \big(rev_s(v) - exp_s(v) \big)}{|C_t|} = \frac{1}{|V_t|} \ .$$
    
    That is, the difference between the diluted balance of each agent and its diluted cash-flow is an equal share of the currency value. 
\end{definition}

Intuitively, a just currency grants an equal share of the currency to every community member, regardless of their initial endowments, while allowing them to do with their share as they please. This results in a socially just allocation of the currency, which is offset from equality only by voluntary trade.  

\begin{observation}[Equal Birth Grant]\label{ob: birth}
    Consider a currency community history where each agent receives a fixed number of coins when it joins the community. Formally, $b_0(v) = x >0$ for all $v\in V_0$ and
    
    $$m_t(v) = \begin{cases} 
    x & \mbox{if } v\in V_t \setminus V_{t-1} \\ 
    0 & \mbox{else } \end{cases}\ .$$
    Such an \textit{equal birth grant regime} is just, as it satisfies 
    
    $$ \frac{b_t(v) - \sum_{s=1}^{t} \big(rev_s(v) - exp_s(v) \big)}{|C_t|} 
    = \frac{b_0(v) + \sum_{s=1}^t m_t(v)}{|C_t|} 
    = \frac{x}{x\cdot |V_t|} = \frac{1}{|V_t|}\ .$$
\end{observation}

Next, we define a relaxed notion of distributed justice.

\begin{definition}[Asymptotic Justice] \label{def: assimptotic justice}
    A currency community history is said to be \textit{asymptotically just}, if 
    
    $$\lim_t \left(\frac{b_t(v)}{|C_t|} - \frac{\sum_{s=1}^{t} \big(rev_s(v) - exp_s(v) \big)}{|C_t|} \right) 
    = \frac{1}{|V|}\ .$$
    
    That is, the difference between the diluted balance of each agent and its accumulative diluted cash-flows converges -- as time advances -- to an equal share of the currency's equity.  
\end{definition}

Intuitively, Definition \ref{def: assimptotic justice} aims to capture justice ``in the limit''. We note that Definition \ref{def: assimptotic justice} is weaker then Definition \ref{def: dist. justice}, that is, a currency community that satisfies distributive justice is also asymptotically just. 

\begin{remark}
    Importantly, we note that both Definitions \ref{def: dist. justice} and \ref{def: assimptotic justice} heavily rely on the currency history being monotone (see Definition \ref{def: community history}). A formal definition of justice in the (very realistic) case of non-monotone histories, as well as the means for achieving it in a setting where agents may die or depart from a community,  would be more subtle. In this paper we refrain from these questions, which include community taxes and inheritance issues, and leave them for future research.
\end{remark}

As demonstrated in Observation \ref{ob: birth}, coin minting may serve as means to achieve distributive justice. In the context of asymptotic justice, we discuss a natural minting regime, termed \textit{egalitarian minting regime}, in which each agents obtain equal income in the form of new coins minted periodically.


\begin{definition}[Egalitarian Minting]\label{def: uniform UBI}
    A currency community history is said to employ \textit{egalitarian minting}, if at every step every agent mints the same amount of coins. Formally,
    $$m_t(v) = \frac{|C_t\setminus C_{t-1}|}{|V_t|}$$ 
    for every $t>0$ and $v\in V_t$.
    %
\end{definition}

Note that egalitarian minting might be realized using a simple digital social contract, as demonstrated by Cardelli et al.~\cite{dsc}. 
The following lemma specifies sufficient conditions under which egalitarian minting is asymptotically just.

\begin{proposition}\label{lem: ubi}
    A currency community history that employs egalitarian minting with $|C_t| \xrightarrow{}\infty$ and $|V| = N < \infty$ is asymptotically just.     %
\end{proposition}

\begin{proof}
Fix an agent $v\in V$ that had joined the community at time $t'$, i.e., $v\in V_{t'}\setminus V_{t'-1}$ and fix $t\geq t'$. By Definition \ref{def: uniform UBI}, we have 
\begin{equation*}\label{eq: sum m1}
\sum_{s=1}^t m_s(v) 
= \sum_{s=t'}^t \frac{|C_s\setminus C_{s-1}|}{|V_s|} 
\geq \sum_{s=t'}^t \frac{|C_s\setminus C_{s-1}|}{N} 
= \frac{|C_t| - |C_{t'-1}|}{N}\ .
\end{equation*}

Consider a time step $t''$ with $|V_{t''}| \geq N$ and fix $t\geq t''$. We then have

\begin{align*}\label{eq: sum m2}
    \sum_{s=1}^t m_s(v) &\leq \sum_{s=1}^{t''-1} \frac{|C_s\setminus C_{s-1}|}{|V_s|} 
    + \sum_{s=t''}^{t} \frac{|C_s\setminus C_{s-1}|}{|V_s|}\\
    &\leq t'' \cdot |C_{t''}| + \frac{|C_t| - |C_{t''-1}|}{N}\ .
\end{align*}

As $|C_{t'}|,|C_{t''}|$ are constant and $|C_t| \xrightarrow{}\infty$, it now follows that

$$\frac{\sum_{s=1}^t m_s(v)}{|C_t|} \xrightarrow{} \frac{1}{N}\ .$$

We thus conclude that

$$\lim_t \left( \frac{b_t(v) - \sum_{s=1}^{t} \big(rev_s(v) - exp_s(v) \big)}{|C_t|} \right) 
= \lim_t \left( \frac{b_0(v) + \sum_{s=1}^t m_t(v)}{|C_t|} \right)
= \frac{1}{|V|}\ .$$
The claim follows.
\end{proof}

To summarize, above we showed that a single, sybil-free currency community that employs egalitarian minting is asymptotically just, namely, as time advances, each member indeed approaches being awarded with an equal share of the currency, offset only by its voluntary trades.
This result is a first step towards the goal of the next section, in which we study the economic relationship between several such currency communities.

\section{Currency Networks}\label{section:currencynetworks}

The egalitarian minting currency described in Section~\ref{section:singlecurrency} indeed satisfies equality and distributive justice, however only for a single, sybil-free community.
Recall that our goal in this paper is a digital currency that is not only equal and just but also grassroots, in that it can support the bottom-up formation of multiple currency communities that can interoperate.
Indeed, we envision that many such currency communities may form independently and we wish to analyze conditions under which all agents in a network of such currency communities will jointly enjoy distributive justice.

To study the economic interactions between different currency communities, the novel mathematical structure we study here is a \emph{currency network}. In this section we present a formal model of currency networks; in particular, we show that they extend and generalize the well-established models of debt and credit networks~\cite{ghosh2007mechanism,dandekar2011liquidity,dandekar2015strategic,ramseyer2019constrained}. We then discuss the exchange rates among independent currencies within the network and formally define the notion of distributive justice in this setting.


\begin{definition}[Currency Network]\label{def:currency network}
A \emph{currency network} is a tuple of currency communities $\calCN = \{\calC^1,...,\calC^k\}$,  $\calC^i = (V^i,C^i,h^i)$, with disjoint sets of coins, $C^i \cap C^j = \emptyset$ for every $i,j \in [k]$.
The currency network has agents $V = \bigcup_i V^i$, coins  $C = \bigcup_i C^i$,  and a \textit{network configuration function} $h:C\xrightarrow{}V$ defined by $h|_{C^i} := h^i$.
\end{definition}

\begin{figure*}[t]
\centering
\includegraphics[width=9cm]{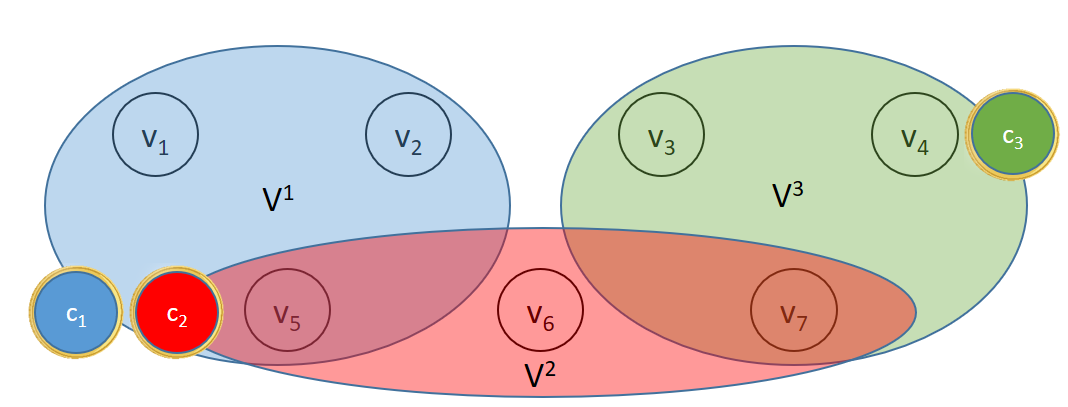}
\caption{A currency network containing $7$ vertices $v_i$, $i \in [7]$ and $3$ communities.
The blue hyperedge on the left ($\{v_1, v_2, v_5\}$) represents the vertices $V^1$ of community $C^1$, the red hyperedge at the bottom represents the vertices $V^2$ of $C^2$, and the green hyperedge on the right represents the vertices $V^3$ of $C^3$.
The agent corresponding to $v_5$ holds the coins $c_1$ of $C^1$ as well as the coin $c_2$ of $C^2$, while the agent corresponding to $v_4$ holds the coin $c_3$ of $C^3$.}
\label{figure:currencynetworkexample}
\end{figure*}

In this model, agents may be members in several communities simultaneously. In order to grasp the network structure, it is useful to think of a currency network as a labeled hypergraph $\calCN = (V,\{V^i\}_{i=1}^k,h)$, where agents $V = \bigcup_i V^i$ are the vertices, and  $\{V^i\}_{i=1}^k$ are the hyperedges, and each vertex $v\in V$ is labeled by the coins it holds from all the communities it is a member of, $h^{-1}(v)$.
See Figure~\ref{figure:currencynetworkexample} for a visual example. We also note that the special case in which all currency communities are of size 2 corresponds to credit networks, where the resulting hypergraph is in fact a graph, as every community is manifested as an edge.

As in a single currency, the fundamental operation in a currency network is a (direct) \emph{payment}, i.e., a transfer of a coin from a payer to a payee (Definition \ref{definition:currency-payment}); However, a payment of a coin of a currency can only be made among two members of the coin's currency community. Still, agents in a currency network may be able to transact with each other via \emph{chain payments}, defined below. 

\begin{definition}[Chain Payment]\label{def: chain pay}
Let $\calCN = \{\calC^1,...,\calC^k\}$ be a currency network, $u,v \in \bigcup_i V^i$. A \emph{chain payment} from $u$ to $v$ is a sequence of direct payments $\calCN_j\xrightarrow{pay(c_j,u_j,u_{j+1)}}\calCN_{j+1}$, from $u_j$ to $u_{j+1}$, $j\in [0,m-1]$, where $u=u_0$, $v = u_m$, and $CN=CN_0$.
\end{definition}

Note that it is not \emph{the same} coin that is transferred among the agents participating in a chain payment.

\begin{observation}
A chain payment from $u$ to $v$ may occur as a contiguous block of transitions if there is a path $p_0=(u_0,u_1,\ldots, u_m)$, $u=u_0$, $u_m = v$, for which each $u_i$ holds a coin acceptable to $u_{i+1}$, $i\in [0,m-1]$.
\end{observation}

By induction on Observation~\ref{ob: rev single currency}, we have that chain payments in currency networks are reversible.

\begin{corollary}[Reversibility in Currency Networks]
If $\calCN$ is a currency network and $\calCN\xrightarrow{pay(c,u,v)}\calCN'$, then
$\calCN'\xrightarrow{pay(c,v,u)}\calCN$.
\end{corollary}

\subsection{Exchange Rates Among Currencies}

Our main aim is to explore the possibility of distributive justice within a currency network. To this end, we first address the issue of exchange rates among the different currencies. For now, we defer the intricate question of the emergence of exchange rates to the next section, and provide a formal definition of exchange rates in this setting, denoting by $\emph{EX}_{ij}$ the amount of coins in $\calC^j$ that may be traded in $\calCN$ for a single coin in $\calC^i$. 

\begin{definition}[Coin Exchange Rates] \label{def: coin ex}
The \textit{coin exchange rates} of a currency network $\calCN = \{\calC^1,...,\calC^k\}$ is given by a matrix $\emph{EX}\in \mathbb{R}^{k\times k}$ that satisfies:
\begin{itemize}
    \item \textbf{Currency fungibility:} $\emph{EX}_{ii} =1$ for all $1\leq i\leq k$.
    \item \textbf{Arbitrage-free trade:} $\emph{EX}_{ij}\cdot \emph{EX}_{jl} = \emph{EX}_{il}$ for all $1\leq i,j,l\leq k$.
\end{itemize}
\end{definition}

That is, coins within the same currency have equal value, and exchanging $c\in\calC^i$ to $\calC^j$ and then to $\calC^l$ yields the same rate as a direct exchange from $\calC^i$ for $\calC^l$. 

\begin{corollary}[Reciprocal rates]
    Let $\emph{EX}\in \mathbb{R}^{k\times k}$ denote a coin exchange matrix of a currency network $\calCN = \{\calC^1,...,\calC^k\}$, then every pair of indices $1\leq i,j\leq k$ satisfy:
    \begin{equation}
    \emph{EX}_{ij} = \frac{1}{\emph{EX}_{ji}}\ . 
    \end{equation}
\end{corollary}

\begin{proof}
    Straightforward from Definition \ref{def: coin ex}.
\end{proof}

Given exchange rates of coins and the total number of coins of each currency, we define the equity of an agent as the  value of its coins as a fraction of the total value of all currencies within the network.

\begin{definition}[Fractional Equity of Agent]\label{def: equity}
    Let $\emph{EX}\in \mathbb{R}^{k\times k}$ denote a coin exchange matrix of a currency network $\calCN = \{\calC^1,...,\calC^k\}$. The \textit{fractional equity of agent} $v\in V$ is given by
    
    $$Eq(v) := \frac{\sum_i b^i(v) \cdot \emph{EX}_{ij}(\calCN)}{\sum_i |C^i| \cdot \emph{EX}_{ij}(\calCN)}\ .$$
\end{definition}


That is, the equity of an agent is the fraction of its assets of the total value of the network, as may be realized in currency $\calC^j$.

\begin{remark}
    We note that Definition \ref{def: equity} is independent of the choice of the index~$j$. To see this, multiply both the nominator and denominator by $\emph{EX}_{jl}(\calCN_t)$ and apply the arbitrage free trade property (see Definition \ref{def: coin ex}). 
\end{remark}

\subsection{Justice Within a Currency Network}

Similarly to the case of a single currency community, our interpretation of distributive justice relies on the dynamics in the network over time. We thus provide the notion of a currency network history.

\begin{definition}[Currency Network History]
A \emph{currency network history} is a sequence of currency networks $\calCN_0, \calCN_1, \calCN_2, \ldots$, $\calCN_t = \{\calC^1_t,...,\calC^k_t\}$, such that $\calC^i_0, \calC^i_1, \calC^i_2, \ldots$, is a currency community history for all $1\leq i \leq k$.
We employ the notation $V:=\bigcup_{i,t} V_t^i$ and $C:=\bigcup_{i,t} C_t^i$ to denote all agents and all coins in the network throughout history.
\end{definition}

In short, a currency network history is nothing but a synchronized set of distinct community histories. As coin exchange rates may vary over time, we apply the notation $\emph{EX}(\calCN_t)$ to differentiate between exchange rates at different time periods throughout history. With the notion of network history at hand, we now extend the notion of distributive justice to a network setting as follows:

\begin{definition}[Distributive Justice in a Network] \label{def: dist. justice network}
    A currency network history $\calCN_0, \calCN_1, \calCN_2, \ldots$ is said to be \textit{just}, if for every $t\geq 0$ and $v\in V_t$:
    
    $$\frac{\sum_i \left[b^i_t(v) - \sum_{s=1}^{t} \big(rev_s^i(v) - exp_s^i(v) \big)\right] \emph{EX}_{ij}(\calCN_t)}{\sum_i |C^i_t| \cdot \emph{EX}_{ij}(\calCN_t)}
    = \frac{1}{|V_t|}\ .$$
\end{definition}

That is, the difference between all assets of an agent and its current cash-flow, exchanged to currency $\calC^j$ and diluted properly, results in each agent's equity being an equal share of the entire currency network's equity, at every time step throughout history. We note that this is a straightforward extension of Definition \ref{def: dist. justice} which corresponds to the special case $k=1$.

Next, we present the notion of asymptotic justice, extended to a network setting.

\begin{definition}[Asymptotic Justice within a Network] \label{def: assimptotic justice network}
    A currency network history $\calCN_0, \calCN_1, \calCN_2, \ldots$ is said to be \textit{asymptotically just}, if for every $v\in V$:
    
    $$ \lim_t \left( \frac{\sum_i \left[b^i_t(v) - \sum_{s=1}^{t} \big(rev_s^i(v) - exp_s^i(v) \big)\right] \emph{EX}_{ij}(\calCN_t)}{\sum_i |C^i_t| \cdot \emph{EX}_{ij}(\calCN_t)} \right)
    = \frac{1}{|V|}\ .$$
\end{definition}

Definitions \ref{def: assimptotic justice network} and \ref{def: dist. justice} for currency networks relate to each other similarly to the way Definitions \ref{def: assimptotic justice} and \ref{def: dist. justice} for a single currency community relate to each other.  Distributive justice in a network requires that the difference between all assets of an agent and its current cash-flow, exchanged to some currency $\calC^j$ and diluted properly, converges to an equal share of the currency network's equity.  Note that Definition \ref{def: assimptotic justice} corresponds to the special case $k=1$.

\section{Justice via Joint Egalitarian Coin Minting}\label{section:justice}

Achieving distributive justice within a currency network requires a joint coin minting regime that is agreeable to all communities in the network. Indeed, the admission of an agent to one community in a just network must affect the distribution of wealth in another, and the exchange rates volatility requires joint efforts in order to maintain distributive justice over time. The joint minting regime required to achieve that is a natural extension of egalitarian coin minting to the network setting.

\begin{definition}[Joint Egalitarian Minting]
    A currency network history is said to employ \textit{joint egalitarian minting} if, at every time step, every agent mints exactly one coin among all currencies in the network:
      Formally, if $\sum_i m_t^i(v) = 1$ for every $t>0$ and $v\in V_t$.
\end{definition}

We demonstrate elsewhere a social contract  for joint egalitarian minting in a currency network~\cite{dsc}. In the following, we explore sufficient conditions under which joint egalitarian minting naturally gives rise to asymptotic justice within all agents participating in multiple currencies within the same currency network. 



\subsection{Myopic Agents}

We begin with the natural question each agent shall ask at each timestep: \emph{Which coin should I mint next?}
Indeed, there are many possibilities. Here we consider a simple answer: \emph{Always mint the highest-valued coin}.

\begin{definition}[Most Valued Coin]\label{def: mvc}
    Let $\calCN = \{\calC^1,...,\calC^k\}$ be a currency network with coin exchange rates $\emph{EX}\in \mathbb{R}^{k\times k}$. A \textit{most valued coin} in this setting is an index $i$ that maximizes $\emph{EX}_{ij}$ over all indices $1\leq j\leq k$.
    Given an agent $v\in V$, a \textit{most valued $v$-coin} is an index $i$ with $v\in V^i$ that maximizes $\emph{EX}_{ij}$ over all indices $1\leq j\leq k$.
\end{definition}

The next definition formalizes the notion of myopic behaviour under egalitarian  minting in a network.

\begin{definition}[Myopic Agents]
Let $\calCN_1, \calCN_2, ...$ be a network history that employs joint egalitarian minting. We say that the agents in the network are \emph{myopic} if in every time step $t$, every agent $v\in V_t$ mints a most valued $v$-coin (ties are broken arbitrarily).
\end{definition}


\subsection{Where do Exchange Rates Come From?}

The relations and interactions among the currencies within a network are inherent to the currency network setting. In the following, we present a conceptual and mathematical framework for the analysis of these interactions which result in exchange rates among the different currencies. We reason that any relation among independent currencies is based upon what the currencies represent, namely actual commodities (e.g., goods and services) that may be purchased from agents that accept these currencies as payment. Specifically, our analysis focuses on the exchange rates that emerge at equilibrium, with respect to individual preferences over these underlying commodities. Note that the commodities are not represented explicitly  in our model; we assume their existence solely to induce preferences on currencies, which we then take into account.

Formally, given a currency network $\calCN = \{\calC^1,...,\calC^k\}$, it will be convenient to view the balances of all agents as a matrix $b\in \mathbb{R}^{n\times k}$, where $b^i(v)$ is the balance of agent $v\in V$ in currency $\calC^i$ ($i$-balance, for short). We denote the \textit{diluted balances} by $\widetilde{b}^i(v) = \frac{b^i(v)}{|C^i|}$, and assume that every agent $v$ has a preference relation $\preceq_v$ over \textit{diluted portfolios} $\widetilde{b}(v) = \left( \widetilde{b}^i(v) \right)_{1\leq i\leq k} \in [0,1]^k$, a vector that corresponds to a fractional ownership in each currency in the network.

This setting is generally known as a \emph{pure exchange economy} (see, e.g.,~\cite{lucas1978asset,moore2007pure,varian1992microeconomic}). We follow standard practice and assume that the preferences of agent $v$ are expressed via a convex, continuous, and monotone linear order over $[0,1]^k$. Given an initial endowment $\widetilde{b}\in [0,1]^{n\times k}$, and assuming that agents may freely trade currencies with each other, the standard solution concept in this model is a competitive equilibrium $\widetilde{b}^*$ wrt.\ the preferences $\{\preceq_v\}_{v\in V}$ that Pareto dominates~$\widetilde{b}$. Importantly, a competitive equilibrium establishes not only an allocation (which is reflected in the balances), but also \textit{marginal rates of substitution} among currencies~\cite{pindyck2005microeconomics}: A matrix $\emph{MRS}\in \mathbb{R}^{k\times k}$ where $\emph{MRS}_{ij}$ denotes the quantity of the currency $\calC^j$ that an agent can exchange for one (infinitesimal) unit of currency $\calC^i$ while maintaining the same level of utility under the equilibrium~$\widetilde{b}^*$.

The normalization of the marginal rates of substitution among currencies by the currency volumes, naturally gives rise to exchange rates among coins within these currencies. As these rates are induced by individual preferences, we term them \textit{preferences-based} exchange rates, formally defined below.

\begin{definition}[Preferences-based rates] \label{def: pref coin ex}
Let $\calCN^*$ be a currency network in which the diluted balances matrix $\widetilde{b}^*$ form an equilibrium under agents' preferences over the currencies. The preferences-based rates between coins in $\calC^i$ and $\calC^j$ is given by 
\begin{equation}\label{eq: coin exchange}
    \emph{EX}_{ij}:= \emph{MRS}_{ij} \cdot \frac{|C^j|}{|C^i|}\ .
\end{equation}
\end{definition}

\begin{remark}\label{remark:remark}
Note the difference between the marginal rate of substitution among currencies (denoted by $\emph{MRS}$), which relates the effective values of the two economies underlying the two compared currencies, and the exchange rate between coins (denoted by $\emph{EX}$). In essence, preferences-based coin exchange rates ($\emph{EX}$) are the currency rates ($\emph{MRS}$), normalized by the number of coins in circulation.
\end{remark}

The following observation asserts that preferences-based rate are valid coin exchange rates as specified in Definition \ref{def: coin ex}. 

\begin{observation}
    Preferences-based rates satisfy currency fungibility and arbitrage free trade.
\end{observation}

\begin{proof}
    As marginal rates of substitution arise in equilibrium, these rates must satisfy both $\emph{MRS}_{ii} =1$ and $\emph{MRS}_{ij} \cdot \emph{MRS}_{jl} = \emph{MRS}_{il}$, or else agents would benefit from further trade. Applying Definition \ref{def: pref coin ex} to these equations completes the proof.
\end{proof}

A key merit of using coins as a medium of exchange (rather than direct trade in fractions of currencies) lies in the degree of freedom manifested in currency volumes, as an increase in money supply causes inflation~\cite{eumoney}. Put simply, if more coins are issued for a single currency, this linearly impact the exchange rate of this currency with other currencies. Roughly speaking, our general approach builds upon the observation that agent choices in coin minting affect and control the fractions $\frac{|C^j|}{|C^i|}$, which in turn affect the coin exchange rates.

We say that the volumes of all currencies are in perfect balance if the ratio between the number of coins of any two currencies exactly equals the difference in the marginal rate of substitution among them in equilibrium. We claim next that if the volumes of a pair of currencies is in perfect balance then a fixed 1:1 coin exchange rate follows.

\begin{observation}\label{ob: centralized}
    Let $\calCN^*$ be a currency network in which the diluted balances matrix $\widetilde{b}^*$ forms an equilibrium under agents' preferences, and let $\emph{EX}$ denote preferences-based coin exchange rates. Then, if two currencies $\calC^i,\calC^j$ satisfy $\frac{|C^i|}{|C^j|} =\emph{MRS}_{ij}$, it follows that $\emph{EX}_{ij}=1$.
\end{observation}

\begin{proof}
    Straightforward from Definition \ref{def: coin ex}.
\end{proof}

Finally, we now aim to extend the notion of exchange rates to all configurations in network history. To do so, we rely on the tendency of reaching equilibria wrt.\ agents' preferences via voluntary mutual trade. Indeed, not all configurations throughout history necessarily form an equilibrium: in particular, it might take several time steps for agents to perform all profitable coin trades and arbitrages. We thus define an efficient history as such that gives rise to  equilibria infinitely often.

\begin{definition}[Efficient History]
Let $\calCN_0,\calCN_1,\calCN_2,\ldots$ be a currency network history with agents' individual preferences over its currencies. Such network history is said to be \textit{efficient} if there exists an (infinite) subsequence $t_1 < t_2 < t_3 \ldots$ such that $\calCN_{t_i}$ is in equilibrium wrt.\ to these preferences. 
\end{definition}

Following that line, we now extend the notion of marginal rates of substitution (and consequently, also preferences-based rates) to all time periods (possibly excluding a finite prefix) by defining the rate at time $t$ as the exchange rate at $t^*$, where $t^*$ is the most recent equilibrium that precedes~$t$. That is, we assume constant rates that are updated occasionally whenever the network reaches equilibrium.

\subsection{Sufficient Conditions for Asymptotic Justice}

Following Observation \ref{ob: centralized}, our aim is to establish 1:1 exchange rates by reaching perfect balance among currency volumes. Our approach builds upon on the dynamics of the trade within the network, as reflected in the network's history. While individual preferences may potentially vary in time, in the following we consider the simple scenario of \textit{fixed agents' preferences}, where $\{\preceq_v\}_{v\in V}$ is fixed eventually, namely after some finite prefix of the currency history in which it may fluctuate.

With the above notions at hand, we can now state our main theorem:

\begin{theorem}\label{thm: main}
    Let $\calCN_0,\calCN_1,\calCN_2,\ldots$ be a currency network history with 2 communities $\calCN_t = (\calC^1_t,\calC^2_t)$ that employs joint egalitarian coin minting. Assume:
    \begin{itemize}
        \item Fixed agents' preferences over the currencies.
        \item Preference-based coin exchange rates.
        \item An efficient network history.
        \item Myopic agents.
    \end{itemize}
    Then, if it holds that
    $$\frac{|V^1 \setminus V^2|}{|V^2|} \leq \lim_t \emph{MRS}_{12}(\calCN_t) 
    \leq \frac{|V^1|}{|V^2\setminus V^1|}\ ,$$
    then the network history is asymptotically just. Furthermore, it also follows that 
    $$\lim_t \emph{EX}_{12}(\calCN_t) = 1\ .$$
\end{theorem}


The proof follows the observation that the agents in the intersection $V^1 \cap V^2$ are the only agents that can choose which coin to mint, and, with myopic joint egalitarian minting, they would choose the more valuable coin; thus, if there are relatively enough agents in the intersection, then, together, they would mint enough coins to set the coin exchange rate right, and asymptotic justice then follows. 

\begin{proof}
    Assuming myopic agents, the number of $\calC^1$-coins minted at each time step $t$, equals $|V^1_t \setminus V^2_t| + \textbf{1}_{\emph{EX}_{12}(\calCN_t) \geq 1} \cdot|V^1_t \cap V^2_t|$, where $\textbf{1}_{\emph{EX}_{12}(\calCN_t) \geq 1}$ is an indicator function that equals $1$ iff $\calC^1$-coins are more valuable then $\calC^2$-coins at time $t$. 
    Consequently, 
    \begin{align*}
        \frac{|C_t^1|}{t} &= \frac{|C_0^1| + \sum_{s=1}^t \left(|V^1_s \setminus V^2_s| + \textbf{1}_{\emph{EX}_{12}(\calCN_s) \geq 1} \cdot|V^1_s \cap V^2_s| \right)}{t}\\ &\rightarrow |V^1 \setminus V^2| + \frac{a_t}{t}\cdot|V^1 \cap V^2|\ , 
    \end{align*}
    where $a_t:= \#\{1\leq s\leq t : \emph{EX}_{12}(\calCN_t) \geq 1\}$ denotes the number of time steps until $t$ in which $\calC^1$-coins are more valuable then $\calC^2$-coins. Similarly, we have $$\frac{|C_t^2|}{t} \rightarrow |V^2 \setminus V^1| + (1-\frac{a_t}{t})\cdot|V^2 \cap V^1|\ .$$ 
    We conclude that 
    \begin{equation} \label{eq: c1/c2}
        \lim_t \frac{|C_t^1|}{|C_t^2|} = \frac{|V^1 \setminus V^2| + \frac{a_t}{t}\cdot|V^1 \cap V^2| }{|V^2 \setminus V^1| + (1-\frac{a_t}{t})\cdot|V^1 \cap V^2|}\ .
    \end{equation}
  
As $\frac{|V^1 \setminus V^2|}{|V^2|} \leq \lim_t \emph{MRS}_{12}(\calCN_t) \leq \frac{|V^1|}{|V^2\setminus V^1|}$, it follows that there exists a unique $0\leq x\leq 1$ for which
\begin{equation} \label{eq: lim MRS}
    \lim_t \emph{MRS}_{12}(\calCN_t) = \frac{|V^1 \setminus V^2| + x\cdot|V^1 \cap V^2| }{|V^2 \setminus V^1| + (1-x)\cdot|V^1 \cap V^2|}\ .
\end{equation}

Now, for sufficiently large $t$, if $\frac{a_t}{t} < x$, it follows from Equations \ref{eq: c1/c2},\ref{eq: lim MRS} that $\frac{|C_t^1|}{|C_t^2|} < \emph{MRS}_{12}(\calCN_t)$, hence, 
$$\emph{EX}_{12}(\calCN_t) = \emph{MRS}_{12}(\calCN_t) \cdot \frac{|C^2_t|}{|C^1_t|} > 1\ .$$
    
That is, $\calC^1$-coins are more valuable then $\calC^2$-coins, thus $a_{t+1} = a_t +1$ and $\frac{a_{t+1}}{t+1} > \frac{a_{t}}{t}$. Similarly, $\frac{a_t}{t} > x$ corresponds to time steps where $\calC^2$-coins are more valuable, hence $\frac{a_{t+1}}{t+1} < \frac{a_{t}}{t}$. 

We conclude that for sufficiently large $t$, $\frac{a_t}{t}$ is monotonically increasing when below $x$ and monotonically decreasing above $x$. As $|\frac{a_{t+1}}{t+1} - \frac{a_{t}}{t}| \xrightarrow[]{} 0$, we conclude that this sequence converges to $x$. It follows that $\lim_t \frac{|C_t^1|}{|C_t^2|} = \lim_t \emph{MRS}_{12}(\calCN_t)$, and therefore 
\begin{align*}
    \lim_t \emph{EX}_{12}(\calCN_t) &= \lim_t \emph{MRS}_{12}(\calCN_t) \cdot \lim_t \frac{|C^2_t|}{|C^1_t|}\\
    &= \lim_t \emph{MRS}_{12}(\calCN_t) \cdot \frac{1}{\lim_t \emph{MRS}_{12}(\calCN_t)} = 1\ .
    \end{align*}
    
In order to establish asymptotic justice, it is enough to note that for sufficiently large $t$: (1) The initial endowment of each agent $v$ (or the exact time of joining each community) is negligible, and (2) Approximate 1:1 rates hold ($\emph{EX}_{12}(\calCN_t) \sim 1$). It follows that 
\begin{align*}
    \frac{\sum_i \left[b^i_t(v) - \sum_{s=1}^{t} \big(rev_s^i(v) - exp_s^i(v) \big)\right] \emph{EX}_{12}(\calCN_t)}{\sum_i |C^i_t| \cdot \emph{EX}_{12}(\calCN_t)} 
    &\sim \frac{const + t}{|C^1_t| + |C^2_t|}\\
    &\xrightarrow{}  \frac{const + t}{t|V|}\\
    &= \frac{1}{|V|}\ ,
    \end{align*}
    which completes the proof.
\end{proof}

\subsection{Examples}

We provide several examples demonstrating the analysis described above.

\begin{example}[Two disjoint communities]
Let $V^1$ and $V^2$ be two communities in some History. If $V^1$ and $V^2$ are disjoint, that is, if $V^1 \cap V^2 = \emptyset$, then the premise of Theorem \ref{thm: main} boils down to $\emph{MRS}_{12}(\widetilde{b}) = \frac{|V^1|}{|V^2|}$, implying that asymptotic justice holds and coin exchange rate approaches 1:1, provided that the relations between the cardinality of the communities are in perfect balance with their \emph{MRS}. This is exactly because each agent will mint a coin of their own currency, thus, in particular, the agents of the community with the higher productivity will ``dilute'' their currency ``exactly'' faster. 

  
%
\end{example}

\begin{example}[Two communities with full intersection]
Conversely, in the case of full intersection, where $V^1 = V^2$, the premise of Theorem \ref{thm: main} boils down to $0 \leq \emph{MRS}_{12}(\widetilde{b}) \leq \infty$. That is, 1:1 exchange rates are guaranteed regardless of the \emph{MRS}.

In other words, a single community that employs an egalitarian minting regime wrt. two currencies, always satisfies asymptotic justice and eventually reaches 1:1 exchange rates:
  This is exactly because all agents are free to select which coin to mint,
  thus would always dilute the highest per-coin-valued currency.
\end{example}

\section{Outlook}\label{section:outlook}

Here we analyzed the possibility of a digital currency that realizes \emph{equality} -- there is not a single entity controlling the currency but all genuine agents equally control the system;
\emph{distributive justice} -- all genuine agents (that is, not including sybils) enjoy an equal share of the value of the digital currency;
and \emph{grassroots} -- several independent communities may freely trade while satisfying joint distributive justice.
Indeed, as we envision bottom-up growth of communities, our analysis, modeled via currency networks, paves the way for interoperability and offers the possibility of equality and justice at scale. 

In particular, our main result shows that joint egalitarian coin minting (that is possible to implement using digital social contracts~\cite{dsc} and in which each agent shall mint only a single coin in each timestep) indeed may lead to pairwise 1:1 exchange rates and thus to joint distributive justice among genuine identifiers~\cite{gid} on currency networks satisfying certain conditions, most importantly sufficient intersections between different currency communities.

Next we discuss some future research directions.

\subsection{Other Regimes}

We analyzed joint egalitarian minting with myopic agents.
Here we mention other possibilities:

\begin{itemize}

    \item \textbf{Egocentric minting:} 
    Here, every agent mints the coin that maximizes her private preferences. (Note that this coin depends both on the agent preferences and on the global exchange rate between coins.)
    
    \item \textbf{Strategic minting:} 
    Here, agents are rational and sophisticated, in that each agent may mint the coin that maximizes its private preferences, taking other agent choices into account.

    \item \textbf{Defensive minting:} 
    Here, in each iteration, each agent mints the coin that it currently has the least among all currencies it is a member of. (This regime can be specified and thus enforced on its parties via a digital social contract.)

\end{itemize}

We leave a detailed study of such possibilities for future work. In particular, studying -- analytically or via computer simulations -- which of these possibilities give rise to 1:1 exchange rate, and what is the rate of convergence, are natural future research directions.

In particular, issues of liquidity in such networks, which could be the main motivation for community merges, shall be studied, as well as the extension of Theorem \ref{thm: main} to networks with more than 2 communities.

Most importantly is the integration of the two approaches - achieving sybil-resilient growth~\cite{csr,gid} of a currency community and  a currency network, using the notion of joint egalitarian coin minting developed
here.

\section*{Acknowledgements}

Ehud Shapiro is the Incumbent of The Harry Weinrebe Professorial Chair of Computer Science and Biology.
We thank the generous support of the Braginsky Center for the Interface between Science and the Humanities.
Nimrod Talmon was supported by the Israel Science Foundation (ISF; Grant No. 630/19).

\bibliographystyle{plain}
\bibliography{bib}

\end{document}